\documentclass[%
 reprint,
nofootinbib,
 amsmath,amssymb,
 aps,
]{revtex4-2}

\usepackage[utf8]{inputenc}
\usepackage{array}
\usepackage{amsmath}
\usepackage{slashed}
\usepackage{amsfonts}
\usepackage{blindtext}
\usepackage[pdftex]{graphicx}
\usepackage[all,cmtip]{xy}
\usepackage[english]{babel}
\usepackage[labelfont=bf, justification=justified]{caption}
\usepackage{verbatim,upgreek}
\usepackage{array}
\usepackage{multirow}
\usepackage{amsfonts}
\usepackage{slashed} 
\usepackage{hyperref}
\usepackage{leading}
\usepackage{scalefnt}
\usepackage[dvipsnames]{xcolor}
\hypersetup{colorlinks=true,linkbordercolor=Blue,linkcolor=Blue, citecolor=Blue}
\usepackage{subcaption}
\usepackage{graphicx}
\usepackage{subeqnarray}
\usepackage{setspace}
\usepackage{indentfirst} 
\usepackage{gensymb}
\newsavebox\myboxA
\newsavebox\myboxB
\newlength\mylenA
\makeatletter
\newcommand{\fmarki}{*}
\newcommand{\fmarkii}{\ensuremath{\dagger}}
\newcommand{\fmarkiii}{\ensuremath{\ddagger}}
\newcommand{\fmarkiv}{\ensuremath{\mathsection}}
\newcommand{\fmarkv}{\ensuremath{\mathparagraph}}
\newcommand{\fmarkvi}{\ensuremath{\|}}
\newcommand{\fmarkvii}{**}
\newcommand{\fmarkviii}{\ensuremath{\dagger\dagger}}
\newcommand{\fmarkix}{\ensuremath{\ddagger\ddagger}}
\newcommand{\fmarkx}{\ensuremath{\mathsection \mathsection}}
\newcommand{\fmarkxi}{\ensuremath{\mathparagraph \mathparagraph}}
                
\def\@fnsymbol#1{{\ifcase#1\or \fmarki\or \fmarkii\or \fmarkiii\or \fmarkiv\or \fmarkv\or \fmarkvi\or \fmarkvii\or \fmarkviii\or \fmarkix \or \fmarkx \or \fmarkxi  \else\@ctrerr\fi}}
\makeatother

\begin{document}

\title[Toward a search for axion-like particles at the LNLS]{Toward a search for axion-like particles at the LNLS}

\author{L. Angel$^{1,2}$}\email{lucia.correa.717@ufrn.edu.br}

\author{P. Arias$^3$}\email{paola.arias.r@usach.cl}

\author{C. O. Dib$^4$}\email{claudio.dib@usm.cl}

\author{A. S. de Jesus$^{1,2}$}\email{alvarosdj@ufrn.edu.br}

\author{S. Kuleshov$^{5,6}$}\email{sergey.kuleshov@unab.cl}

\author{V. Kozhuharov$^7$}\email{venelin.kozhuharov@cern.ch}

\author{L. Lin$^8$} \email{liu@lnls.br}

\author{M. Lindner$^9$}\email{lindner@mpi-hd.mpg.de}

\author{F. S. Queiroz$^{1,2,6}$}\email{farinaldo.queiroz@ufrn.br}

\author{R. C. Silva$^{1,2}$}\email{ricardo.rego.115@ufrn.edu.br}

\author{Y. Villamizar$^{10}$}\email{yoxarasv@sprace.org.br}

\affiliation{$^1$ Departamento de F\'isica Te\'orica e Experimental, Universidade Federal do Rio Grande do Norte, 59078-970, Natal, Rio Grande do Norte, Brasil} 

\affiliation{$^2$ International Institute of Physics, Universidade Federal do Rio Grande do Norte, Campus Universit\'ario, Lagoa Nova, Natal, Rio Grande do Norte, 59078-970, Brasil}

\affiliation{$^3$ Departamento de Física, Universidad de Santiago de Chile, Casilla 307, Santiago, Chile}

\affiliation{$^4$ Departamento de Física and CCTVal, Universidad Técnica Federico Santa María, Valparaíso 2340000, Valparaíso, Chile}

\affiliation{$^5$ Center for Theoretical and Experimental Particle Physics, Facultad de Ciencias Exactas, Universidad Andres Bello, Fernandez Concha 700, Santiago, Chile}

\affiliation{$^6$ Millennium Institute for Subatomic Physics at High-Energy Frontier (SAPHIR), Fernandez Concha 700, Santiago, Chile} 

\affiliation{$^7$ Faculty of Physics, Sofia University, 5 J. Bourchier Blvd., 1164, Sofia, Bulgaria}

\affiliation{$^8$ Laborat\'orio Nacional de Luz Síncrotron - LNLS, Caixa Postal 6192, CEP 13084-971,  Campinas, São Paulo, Brazil }

\affiliation{$^9$ Max Planck Institut fur Kernphysik, Saupfercheckweg 1, 69117 Heidelberg, Germany}

\affiliation{$^{10}$ Centro de Ci\^encias Naturais e Humanas, Universidade Federal do ABC, 09210-580, Santo Andr\'e, S\~ao Paulo, Brasil} 

\begin{abstract}
    
Axion-Like Particles (ALPs) appear in several dark sector studies. They have gained increasing attention from the theoretical and experimental community. In this work, we propose the first search for ALPs to be conducted at the Brazilian Synchrotron Light Laboratory (LNLS). In this work, we derive the projected sensitivity of a proposed experiment for the production of ALPs via the channel $e^+ e^- \to a \gamma$. We show that such an experiment could probe ALP masses between $1-55\,\mbox{MeV}$, and ALP-electron couplings down to $g_{aee}=2-6\times10^{-4} \,\mbox{GeV}^{-1}$ depending on the energy beam, thickness of the target, and background assumptions. Therefore, this quest would cover an unexplored region of parameter space for experiments of this kind, constitute a promising probe for dark sectors, and potentially become the first Latin-American dark sector detector.
\end{abstract}


\maketitle

\section{Introduction}\label{sec1}

The Standard Model (SM) of particle physics has successfully described numerous observed phenomena in nature. However, there remain some missing pieces to the puzzle. One common extension to the SM is the inclusion of new neutral bosons, such as vector, scalar, and pseudoscalar particles. Among the pseudoscalar extensions, the axion is one of the most well-known, having been first introduced in the literature in 1977 \cite{PecceiQuinn1, PecceiQuinn2} as a solution to the strong CP problem in QCD. By introducing a new global symmetry, the Peccei-Quinn symmetry, which is spontaneously broken at some high energy scale and explicitly broken due to instanton effects, the mechanism can resolve the Strong CP problem and predict the existence of a new particle, the axion. Alongside axions, new pseudoscalar particles, known as axion-like particles (ALPs), have also emerged as possible extensions to the SM.

Analogously to axions, ALPs have an effective coupling to photons and possibly fermions. However, for ALPs, their coupling constant and mass are independent parameters. Their popularity arises from several aspects: if they are light enough, they can be produced non-thermally in the early universe (e.g., via the misalignment mechanism proposed by Dine and Fischler in 1982 \cite{Dine:1982ah}), without the need for messenger particles, and potentially account for the entire dark matter content observed today within a wide parameter space \cite{Masso:1995tw, Arias:2012az}. Slim ALPs in a higher mass range, with masses below or around MeV, have been extensively sought after at the intensity frontier \cite{Proceedings:2012ulb}.

In the intermediate mass range, from MeV up to a few GeV, ALPs have received increasing attention. Their potential to alleviate the tension observed in measuring the anomalous electron and muon magnetic moments \cite{Bauer_2017, Chang:2000ii, Darme:2020sjf} and as messengers between the SM and invisible (dark matter) sectors \cite{PhysRevD.83.115009,PhysRevD.90.015012} have been recognized. Constraints on this mass range are mainly provided by colliders, such as BaBar \cite{BaBar:2017tiz}, CLEO \cite{CLEO:1994hzy}, LEP \cite{Jaeckel_2016}, LHC, and CLIC \cite{Bauer_2017}, as well as beam dump experiments.

So far, the most studied coupling for ALPs is the one with two photons, which is highly constrained. However, the coupling to fermions still presents an important gap, especially for coupling constants $g_{aee}\lesssim 10^{-2}$~GeV$^{-1}$, for masses in the GeV range. Supernovae (SN) are potential sources of ALPs, and they provide strong constraints on several couplings and mass ranges \cite{Ge:2020zww,Sakstein:2022tby,Mori:2022kkh}. ALPs in the MeV mass range that feature a coupling to photons are mainly produced through the Primakoff effect, and they later decay, producing a gamma ray flux that sets the upper bound $g_{a\gamma \gamma}<10^{-11}$~GeV$^{-1}$ for a mass $m_a\sim 10$~MeV \cite{Jaeckel:2017tud,Calore:2021klc}.

However, if ALPs couple to electrons, they can be produced in the core of a supernova mainly through electron Bremsstrahlung, or if their mass is greater than 10 MeV, through electron-positron annihilation. For sufficiently weak couplings, the ALPs can escape the supernova and decay on their way to Earth. This production mechanism can constrain the parameter space to $g_{aee}>5\times 10^{-7}$~GeV$^{-1}$ for $m_a\sim 120$ MeV \cite{Lucente:2021hbp}. Despite these severe restrictions, it is important to check the aforementioned parameter space with laboratory experiments that have controlled sources and backgrounds, as the astrophysical environments are not fully modeled in some cases, with significant improvements currently underway \cite{Carenza:2019pxu}.

\begin{figure}
    \centering
    \includegraphics[width=\columnwidth]{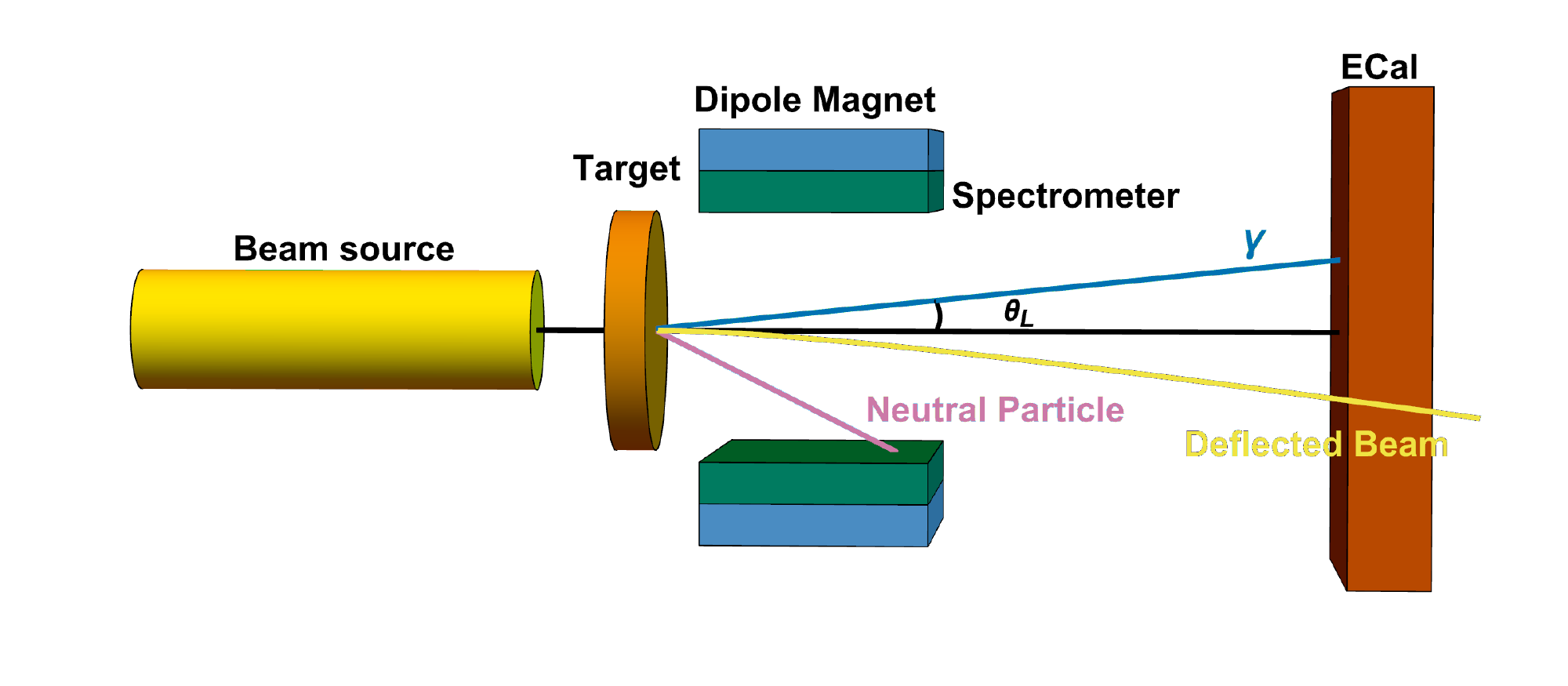}
    \caption{Experimental scheme of the SeDS experiment. In this scheme, we show the main parts of the experiment, such as the beam source in yellow, the diamond target in orange, the dipole magnet in blue, the spectrometer in green, and the electromagnetic calorimeter in vermilion.}
    \label{experiment-scheme}
\end{figure}

In this study, our objective is to develop a theoretical proposal for an ALP search, and evaluate new physics reach of Search for Dark Sector (SeDS), a detector for the dark matter sector \cite{Duarte:2022feb} at the UVX Synchrotron. UVX is a second-generation synchrotron light source that used accelerate electrons to $1.37$~GeV \cite{Lin:1993np,Liu:2010bfa,Lescano:2017vlq}, but it has recently been decommissioned. UVX has been succeeded by Sirius, a fourth-generation storage ring \cite{Rodrigues:2019oej,Alves:2019aei,Liu:2021txs}. Certain subsystems of UVX may be reutilized to develop a new $1-3$~GeV positron accelerator, which would give rise to SeDS, a small-scale fixed target experiment aimed at searching for dark sectors, including ALPs. The design of the proposed detector is illustrated in Fig. \ref{experiment-scheme}.

The production process will occur through electron-positron annihilation processes, specifically via channels of the type $e^+ e^- \to a \gamma$ as illustrated in Fig. \ref{feynman-diagrams}. The SeDS experiment will feature a positron beam consisting of 10 bunches of $10^{9}$ positrons on target (POT) per second, with energies ranging from $1$ to $3 \,\mbox{GeV}$ directed at a diamond target with a thickness of $d=100 - 500 \,\mu\mbox{m}$. It is well-known that with such an intense beam, a portion of the beam will pass through the target. To address this issue, a magnetic dipole will be added to sweep the non-interacting positron beam away from the calorimeters. This dipole magnet will function as a filter, reducing the number of events in the electromagnetic calorimeter (ECal). Furthermore, a spectrometer is expected to be installed between the gap in the dipole magnet to detect possible positrons that emitted would radiate bremsstrahlung photons, suppressing this background source.  
On the other hand, since the photons are not affected by the magnetic dipole, they will hit the calorimeter, be measured, and used to reconstruct the ALP mass using the missing mass technique. This setup is expected to yield accurate and precise measurements of the ALP mass. We will show that SeDS has the potential to significantly contribute to the search for ALPs, and to deepen our understanding of the possible dark sectors.

The missing mass technique is based on 4-momentum conservation, depending only on the initial parameters of the setup and knowledge about the outgoing photon 4-momentum. With the missing mass technique, it is possible to predict the mass of the ALP to be used in the calculations, given by
\begin{eqnarray}
 M^2_{\text{miss}}=(p_{e^-} + p_{\text{beam}}-p_{\gamma})^2,\label{missing-mass-tech}
\end{eqnarray} where $p_{e^-}$, $p_{\text{beam}}$ and $p_{\gamma}$ are the 4-momentum of the target's electrons, positron beam and outgoing photon, respectively. 

The rest of this article is organized as follows. In Section \ref{sec2} we describe the model for the ALP we consider. In Section \ref{sec12} we analyze and discuss the observability and bounds of the ALP. Section \ref{sec13} contains our conclusions. 

\section{\label{sec2} The Model}

As our model, we consider a straightforward case of ALPs that interact only with photons and electrons\footnote{Actually, even if we remove the photon interaction at tree level, a similar term would be induced at loop level \cite{Bauer_2017}.}, resulting in the following lagrangian,
\begin{eqnarray}
    \mathcal{L} \supset \frac{1}{2}(\partial_\mu a)(\partial^\mu a) - \frac{1}{2}m_a a^2 + \frac{1}{4}g_{a\gamma \gamma} a F_{\mu \nu} \Tilde{F}^{\mu \nu}\nonumber \\
    + \frac{g_{aee}}{2}(\partial_\mu a) (\Bar{e}\gamma^\mu \gamma^5 e),
\end{eqnarray} where $F_{\mu \nu}$ is the electromagnetic strength tensor, $\Tilde{F}^{\mu \nu} = \frac{1}{2}\epsilon^{\mu \nu \alpha \beta} F_{\alpha \beta}$ is the dual tensor, $m_a$ is the ALP mass, $g_{a\gamma \gamma}$ is the effective coupling of the ALPs with the photon and $g_{aee}$ being the effective coupling of the axial interaction of the ALPs with the electron, both with dimension of inverse energy.

Since the ALP only interacts with photons and electrons, it has two main decay channels, $a \to \gamma \gamma$ and $a \to e^+ e^-$. The decay widths of these channels are given by,
\begin{eqnarray}
    \Gamma_{a \to \gamma \gamma} &=& \frac{g_{a\gamma \gamma}^2m_a^3}{64 \pi};\\
    \Gamma_{a \to e^+ e^-} &=& \frac{g_{aee}^2}{8\pi} m_e^2 m_a \sqrt{1-\frac{4m_e^2}{m_a^2}}.
    \label{eq:int_gamma}
\end{eqnarray}

\begin{figure}
    \centering
     \begin{subfigure}[t]{0.4\textwidth}
         \centering
         \includegraphics[scale=0.7]{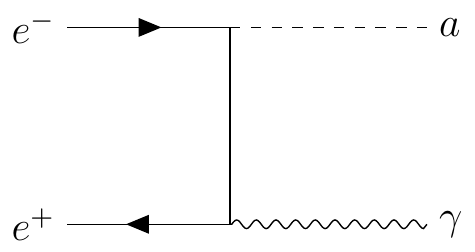}
         \caption{Electron mediated channel.}
         \label{fig:electron channel}
     \end{subfigure}
     \hfill
     \begin{subfigure}[t]{0.4\textwidth}
         \centering
         \includegraphics[scale=0.6]{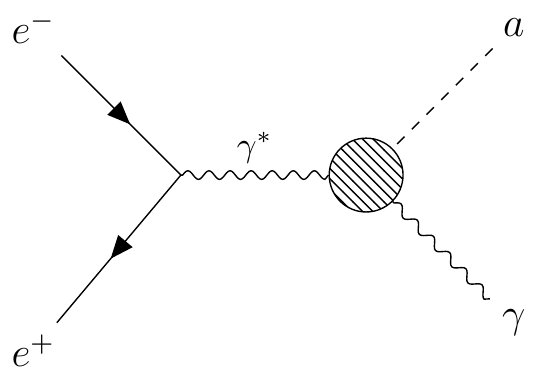}
         \caption{Photon mediated channel.}
         \label{fig:photon channel}
     \end{subfigure}
    \caption{Feynman diagrams of the channels that contribute to the ALP production via electron-positron annihilation into a photon and the ALP ($e^+e^-\to a \gamma$).}
    \label{feynman-diagrams}
\end{figure}

We remind the reader that our proposal is based on the search for ALPs via the process $e^+ e^- \to a \gamma$. There are two different channels that can produce ALPs with this signature, represented by the diagrams in Fig. \ref{feynman-diagrams}. The total cross-section of these channels are given by $\sigma_T = \sigma_{a \gamma} + \sigma_{a e} + \sigma_{int}$, with,
\begin{widetext}
\begin{eqnarray}
    \sigma_{a \gamma} &=& \alpha_{em}g_{a\gamma \gamma}^2\frac{(s+2m_e^2)(s-m_a^2)^3}{24\beta s^4},\\
    \sigma_{a e} &=& \alpha_{em}g_{aee}^2m_e^2\frac{-2m_a^2\beta s +(s^2+m_a^4-4m_a^2m_e^2)\log\frac{1+\beta}{1-\beta}}{2(s-m_a^2)s^2\beta^2},\label{sae}\\
    \sigma_{int} &=& \alpha_{em}g_{a\gamma \gamma}g_{aee}m_e^2 \frac{(s-m_a^2)^2}{2\beta^2 s^3}\log\frac{1+ \beta}{1-\beta},
\end{eqnarray} \end{widetext} where $\sqrt{s}= \sqrt{2m_e(m_e+E_{beam})}$ is the center of mass (CM) energy, $m_e$ is the electron mass, $\beta=\sqrt{1-4m_e^2/s}$ and $\alpha_{em}=e^2/4\pi$ is the electromagnetic fine structure constant. The first equation ($\sigma_{a\gamma}$) is the contribution from the diagram in Fig. \ref{fig:electron channel}, representing the cross-section of the photon-mediated channel as a function of $m_a$, whilst the second equation ($\sigma_{ae}$) is for the diagram in Fig. \ref{fig:photon channel}. The last equation is the interference term between the two contributions.

It is important to note that ALPs can be produced through other channels, such as Bremsstrahlung with the target ($e^+ + N \to e^+ + N + a$) and Primakoff production originating from a secondary photon that collides with the target ($\gamma + N \to N + a$). However, these channels produce a distinct signal from the previously discussed ones, and the missing mass technique described in Eq. (\ref{missing-mass-tech}) cannot be employed in the same manner since there are no photons in the final state. Hence, these channels do not represent a feasible signal for our setup. An important ingredient of dark sector searches is the decay length. One needs to know whether the ALP will decay inside or outside the detector.

\begin{figure*}[ht!]
    \centering
    \begin{subfigure}[t]{0.49\textwidth}
         \centering
         \includegraphics[width=\textwidth]{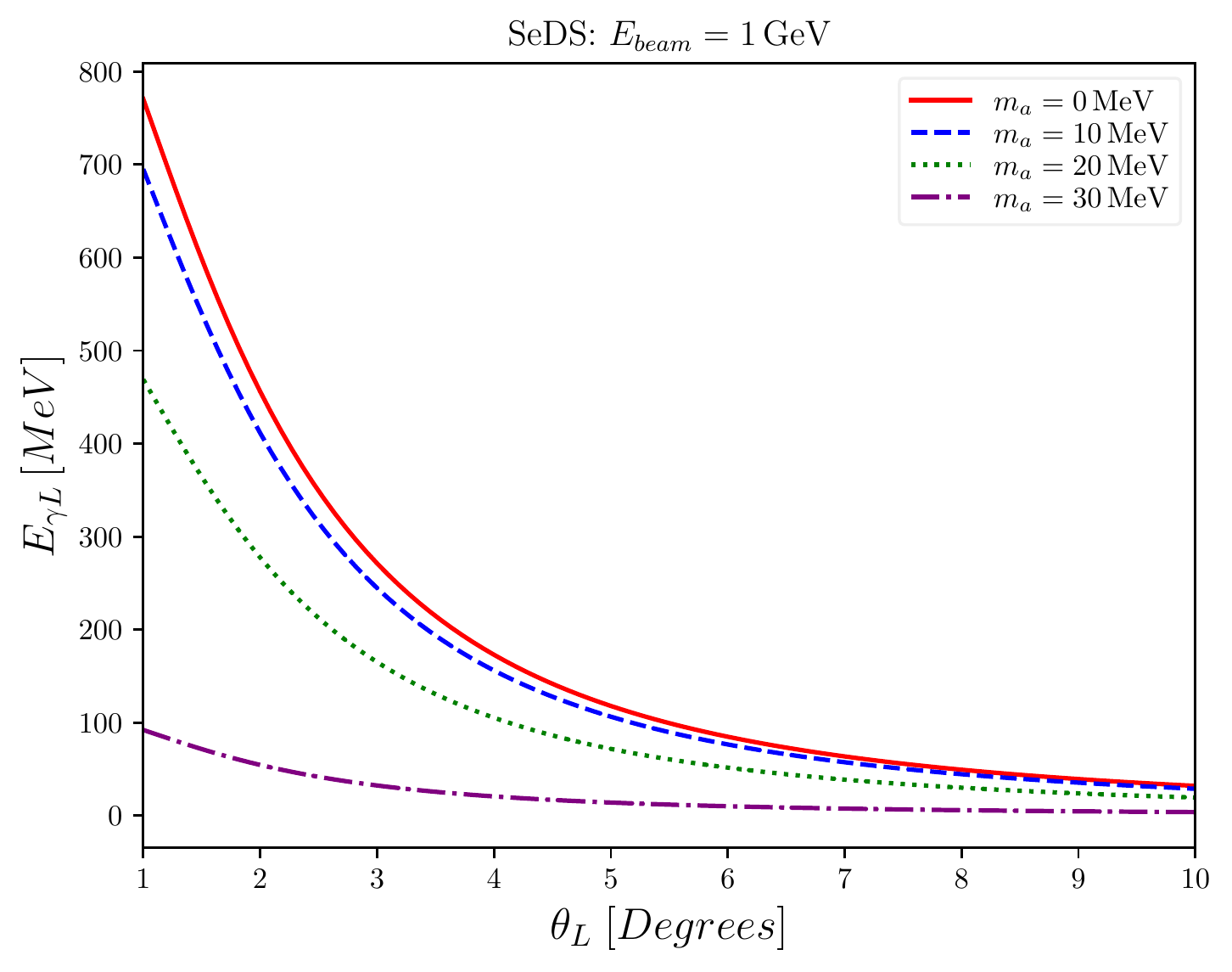}
         \caption{Photon energy for $E_{beam}=1 \, \mbox{GeV}$.}
         \label{fig:photonenergy1gev}
     \end{subfigure}
     \hfill
     \begin{subfigure}[t]{0.49\textwidth}
         \centering
         \includegraphics[width=\textwidth]{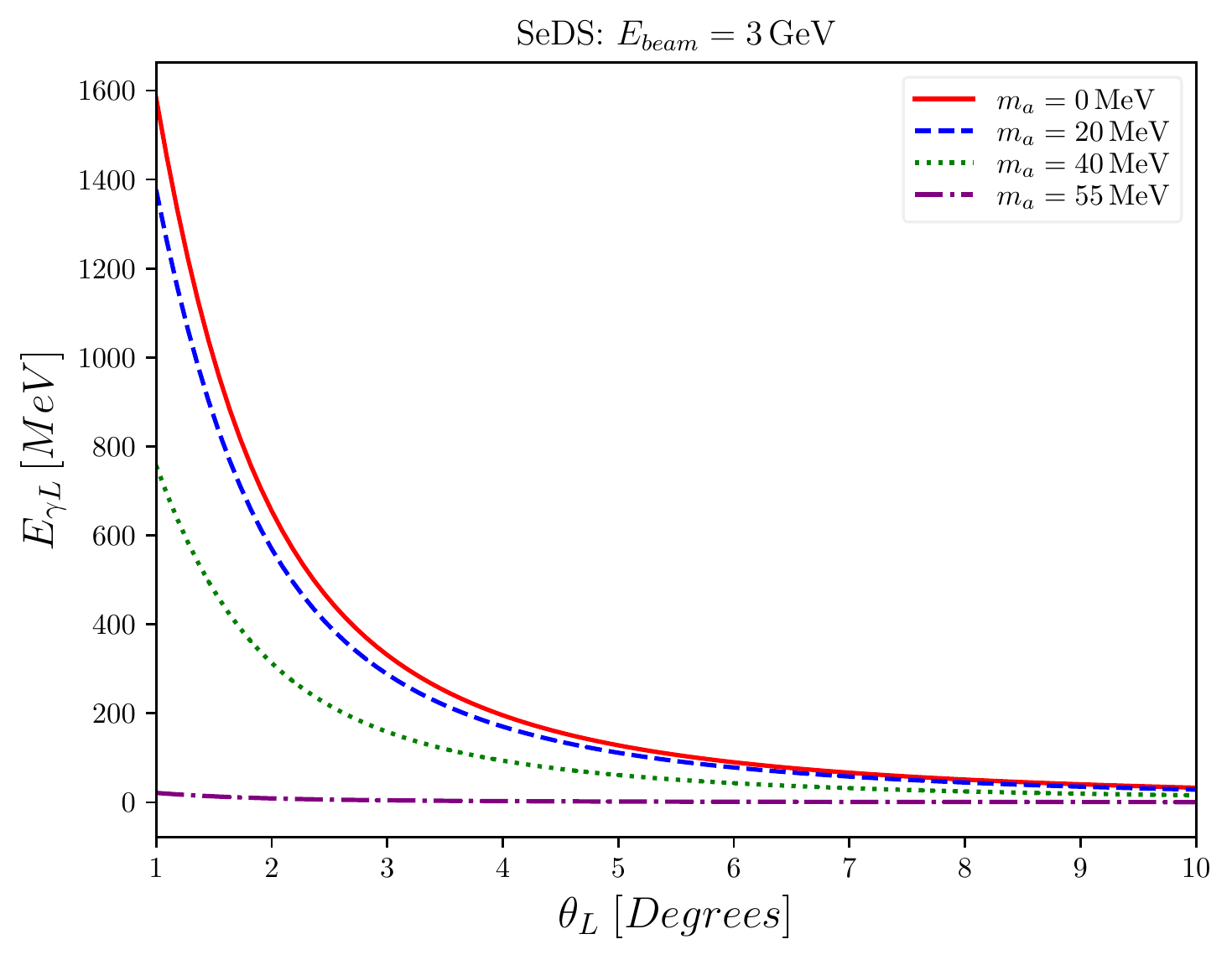}
         \caption{Photon energy for $E_{beam}=3 \, \mbox{GeV}$.}
         \label{fig:photonenergy3gev}
     \end{subfigure}
    \caption{Energy of the photon produced along with the ALP as a function of the opening angle between the photon and the beam axis in the Laboratory frame for different configurations of the ALP mass. As expected, the photon's energy gets smaller as larger masses are assumed.}
    \label{photonenergy}
\end{figure*}

\begin{subsection}{Decay Length}

The first method presented in this work is the so-called invisible search, which is concentrated on the visible counterpart of the signal, i.e., the photon in the final state. However, depending on the detector setup and properties of the ALP, the decay products of the ALP can be measured, if it decays inside the detector. The latter is known as visible search. A key quantity in this case is the decay length, which is found to be,
\begin{equation}
    L_{a}=\gamma \beta c \tau_{a} \approx \frac{E_{a}}{m_{a}\Gamma_{a}},
    \label{Eqdecaylength}
\end{equation} where $\gamma=E_{a}/m_{a}$ is the Lorentz factor, $\tau_{a}=\Gamma_{a}^{-1}$ is the lifetime of the ALP, which is inversely related to the decay width presented in Eq. (\ref{eq:int_gamma}), $\beta = \sqrt{1-4m_e^2/s}$, and $E_{a}$ being the energy of the ALP. In order to determine the decay length of the ALP to assess whether it decays inside or outside the detector, we need to know the energy of the ALP, $E_a$. The energy of the ALP is related to the outgoing photon energy,
\begin{equation}
    E_{a} = \frac{s}{2 m_e} - E_{\gamma L}, \label{Ea-Egamma-relation}
\end{equation} where $E_{\gamma L}$ is the energy of the photon in the laboratory system, which  depends on the opening angle of the photon in relation to the beam axis, and the mass of the ALP as follows,
\begin{equation}
    E_{\gamma L} = \left(1 - \frac{m_{a}^2}{s} \right)m_e\frac{1+ \beta \cos^2\theta_L \sqrt{1+\tan^2\theta_L}}{1-\beta^2\cos^2\theta_L}, \label{EgL}
\end{equation} where $\beta$ is the velocity of the CM in the laboratory frame and $\theta_L$ is the angle between the photon and the beam axis. 

In order to clearly demonstrate how the energy and angle of the photon detected in the calorimeter are influenced by the ALP mass, we present the results from Eq. (\ref{EgL}) for two different beam energies in Fig. \ref{photonenergy} from where it can be inferred that the energy of the outgoing photon increases as the opening angle decreases. Furthermore, in order to accurately determine the mass of the ALP produced in the collision, it is crucial to use a calorimeter with a good energy resolution, especially for photons detected at a high opening angle. 

The energy measurement of an electromagnetic calorimeter is determined by the energy released in the detector material through ionization and excitation processes, which is proportional to the energy of the incident particle. The thickness of the absorber layers in radiation lengths affects the energy resolution of an electromagnetic calorimeter, with a smaller thickness leading to a larger number of detected particles and a better energy resolution \cite{Fabjan:2003aq}. To achieve precise determination of the photon energy and ALP mass, it is important to reduce the thickness of the absorber layers. A Bismuth Germanate (BGO) calorimeter has achieved an energy resolution of $2\%/\sqrt{E{\rm (GeV)}}$ \cite{Frankenthal:2018yvf,Bantes:2015sma}, which is sufficient for our scientific goal. For example, a $100$~MeV photon would yield a 6\% energy resolution. A Lutetium Yttrium Orthosilicate (LYSO) calorimeter one could improve this energy resolution by a factor of two \cite{Anderson:2015gha,Bornheim:2017gql,Saad:2020wav} though.  

Having that in mind, we use the measured $E_{\gamma L}$, combined with knowledge of the beam energy to derive $E_a$, and subsequently determine the decay length of the ALP using Eq. (\ref{Eqdecaylength}) for $E_{beam} = 1\, \mbox{GeV}$  (Fig. \ref{fig:decaylength1gev}) and $E_{beam} = 3 \, \mbox{GeV}$ (Fig. \ref{fig:decaylength3gev}). In Fig. \ref{alpdecaylength}, we present the decay length of the ALP as a function of the photon angle in the laboratory frame $\theta_{L}$, assuming a coupling of the ALP with the electron of $g_{aee} = 10^{-3}\, \mbox{GeV}^{-1}$, and various values of $m_a$ within the region of interest. Notably, for this configuration, the decay length of the ALP is greater than 20 meters, resulting in the ALP being invisible. Therefore, our proposal is designed to detect invisible ALPs. Having established the search for invisible ALPs we will address the production cross section.

\begin{figure*}
    \centering
    \begin{subfigure}[t]{0.49\textwidth}
         \centering
         \includegraphics[width=\textwidth]{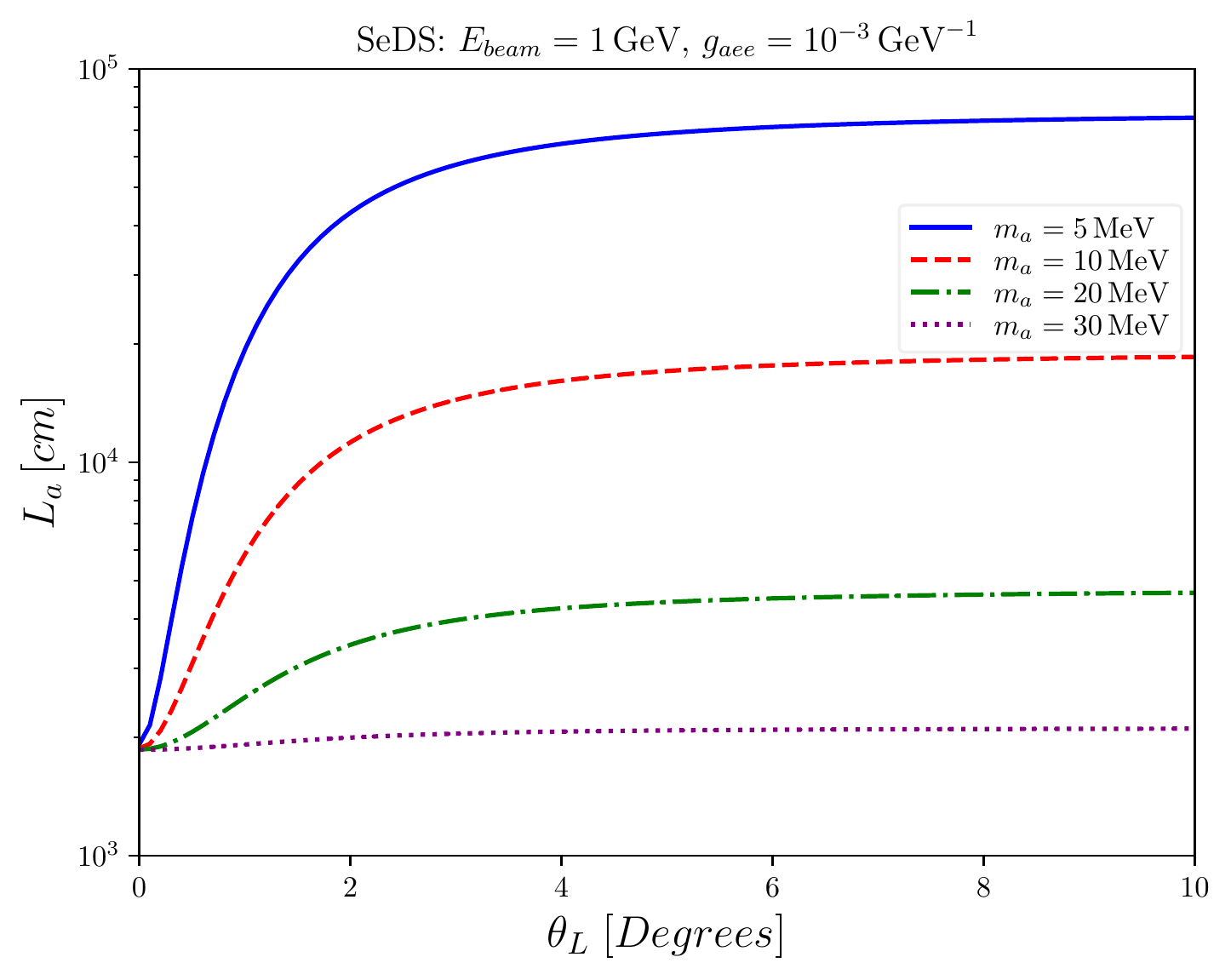}
         \caption{Decay length for $E_{beam}=1 \, \mbox{GeV}$.}
         \label{fig:decaylength1gev}
     \end{subfigure}
     \hfill
     \begin{subfigure}[t]{0.49\textwidth}
         \centering
         \includegraphics[width=\textwidth]{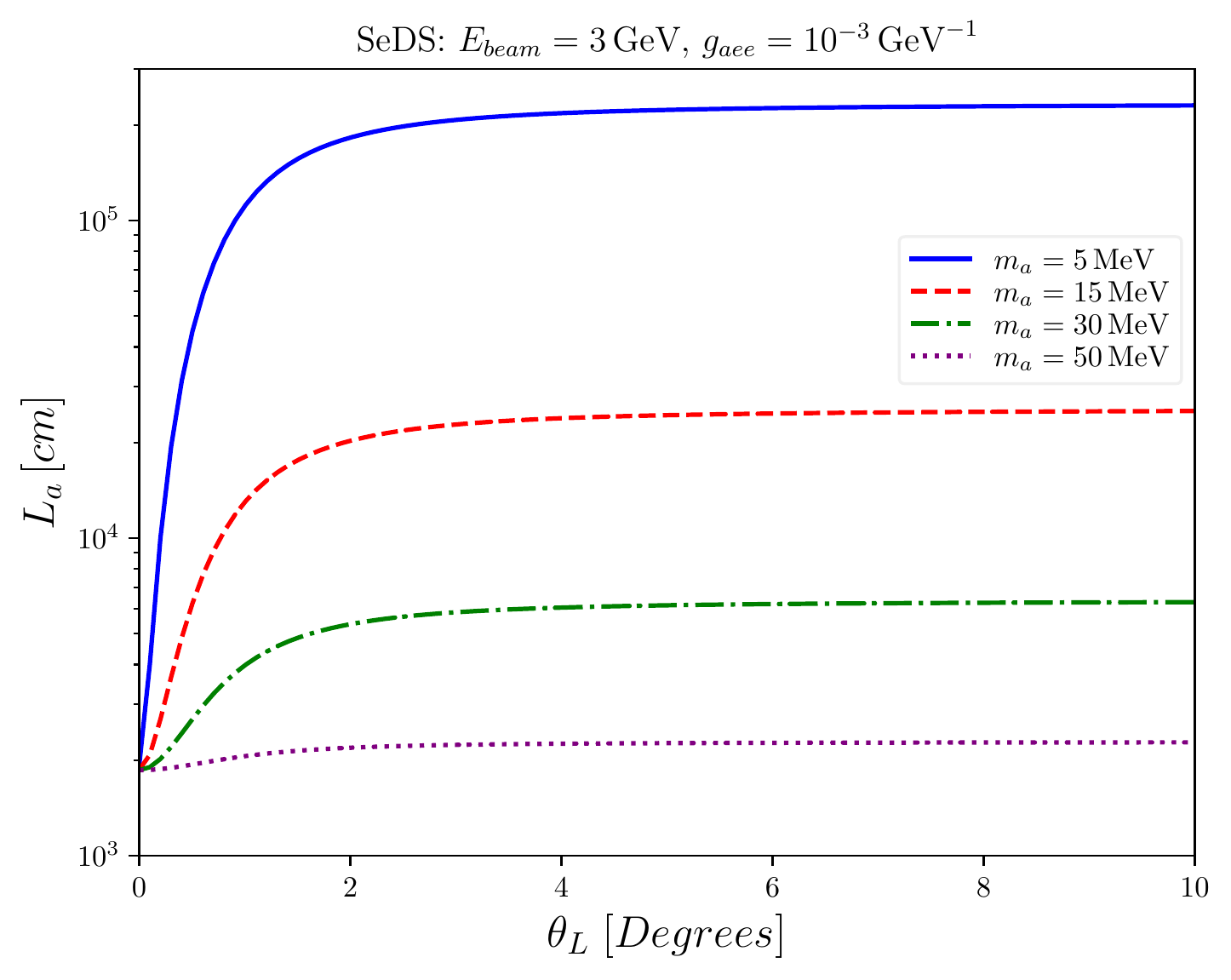}
         \caption{Decay length for $E_{beam}=3 \, \mbox{GeV}$.}
         \label{fig:decaylength3gev}
     \end{subfigure}
    \caption{Decay length of the ALP as a function of the photon angle $\theta_{L}$ in the laboratory frame, taking $g_{aee} = 10^{-3}$ for $E_{beam}=1 \, \mbox{GeV}$ (Fig. \ref{fig:decaylength1gev}) and $E_{beam}=3 \, \mbox{GeV}$ (Fig. \ref{fig:decaylength3gev}). Notably, the ALP will decay outside the detector.}
    \label{alpdecaylength}
\end{figure*}

\end{subsection}

\section{\label{sec12} Discussion}

\begin{figure*}
    \centering
    \includegraphics[width=.6\textwidth]{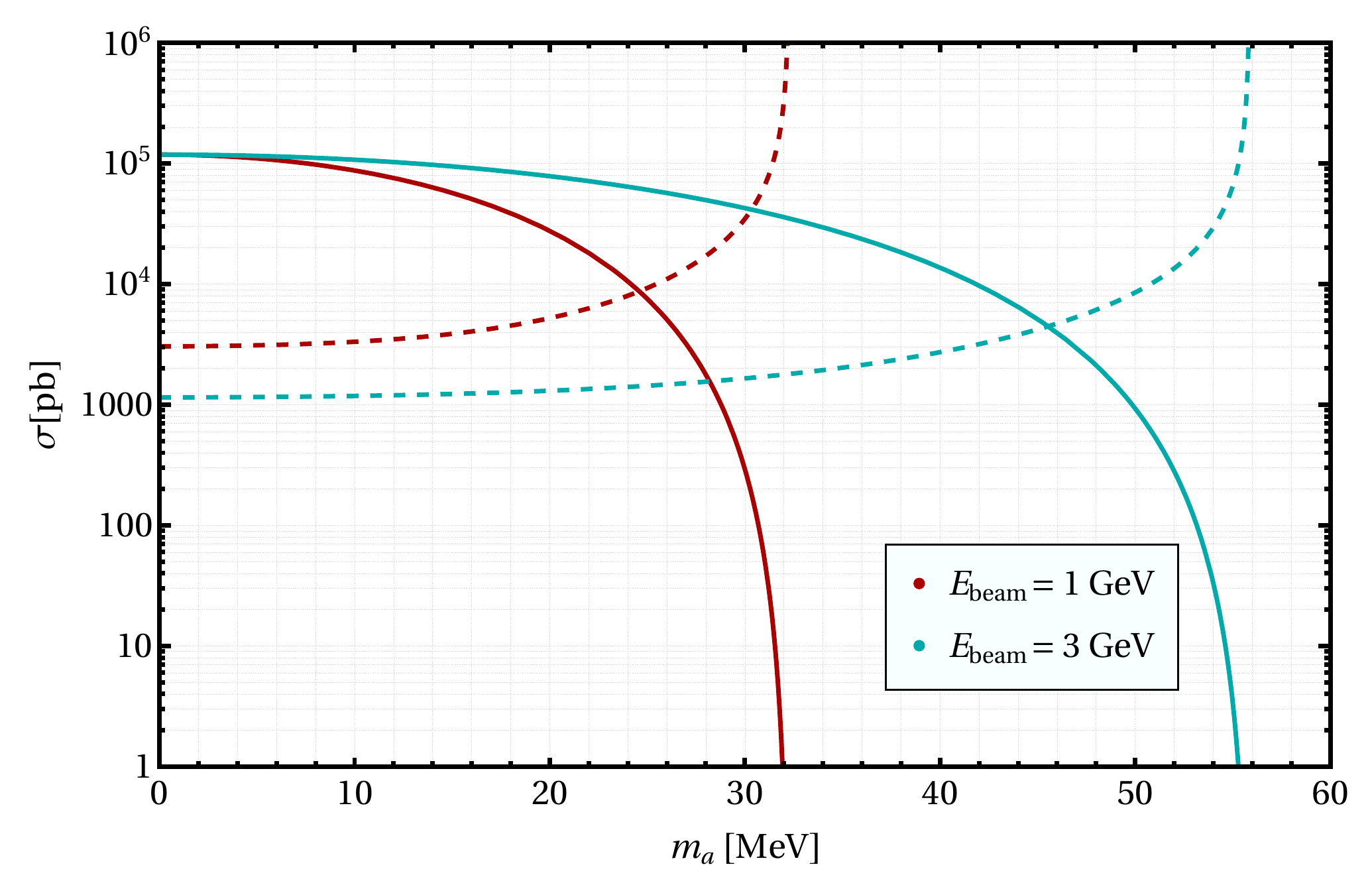}
    \caption{Cross-section of the photon (solid lines) and electron (dashed lines) mediated processes of ALP production as a function of the ALP mass $m_a$. We considered $g_{a\gamma \gamma}=g_{aee}=1 $ GeV$^{-1}$ and beam energies of 1 GeV and 3 GeV. The drop in the cross section occurs due to kinematics for $m_a=32$~MeV and $m_a=55$~MeV, respectively.}
    \label{crosssectionplot}
\end{figure*}

Upon analyzing the total cross-section, it has been observed that the interference term is suppressed by a factor of $m_e^2$. As a result, the contribution of the interference term to the total cross-section is at most 3.5\% compared to the case where no interference is present. Thus, we will ignore it and focus on investigating the two ALPs production cross sections, namely $\sigma_{a\gamma}$ and $\sigma_{ae}$, independently.

The contributions of the photon and electron mediated channels can be seen in Fig. \ref{crosssectionplot}, where we plotted $\sigma_{a\gamma}$ and $\sigma_{ae}$ for the beam energies of $E_{beam}= 1\,\mbox{GeV}$  and $E_{beam}=3\,\mbox{GeV}$. The contribution of the photon (electron) channel is presented in solid (dashed) lines. We assumed $g_{a\gamma \gamma}=g_{aee}=1 $ GeV$^{-1}$. Although we acknowledge that $g_{a\gamma \gamma}=1$ GeV$^{-1}$ is excluded by observations, we aim to emphasize the significance of $g_{a\gamma\gamma}$ in the production of ALPs via $e^+e^-$ annihilations. Furthermore, when $m_a \simeq \sqrt{s}$, we find a divergence at $\sigma_{ae}$, as shown in the denominator of Eq. (\ref{sae}). 

When the couplings have comparable magnitudes, the contribution from the photon-mediated process dominates over most of the parameter space. However, as the mass of the ALP approaches the center of mass energy, the electron-mediated process becomes the dominant channel. The center of mass energy is determined by $\sqrt{s}=\sqrt{2E_{\text {beam}} m_e}$ and is equal to 31.9 and 55.3 MeV for beam energies of 1 and 3 GeV, respectively. Hence, when $g_{a\gamma \gamma} \ll g_{aee}$ the ALP production is governed by $g_{aee}$. Having in mind the restrictive bounds on $g_{a\gamma\gamma}$, we will assume that $g_{aee}$ is sufficiently larger than $g_{a\gamma\gamma}$, hereafter. That said, we will concentrate our work on $g_{aee}$ in what follows.

\subsection{Current and Projected Bounds on the model}

There are several limits on ALPS stemming from beam dumps experiments \cite{Banerjee_2020,PhysRevD.38.3375,PhysRevLett.59.755}, cosmology \cite{Ghosh_2020}, and astrophysical sources \cite{Lucente:2021hbp,Chang_2018}. Most of these constraints arise from astrophysical sources. However, probing such interactions in accelerators would be desirable as it would be subject to smaller systematic uncertainties. In this study, we will focus on the couplings between ALPs and electrons, specifically $g_{aee}$, which can be directly probed using $e^+e^-$ collisions. To assess the relevance of our proposal, we will put our findings into perspective with existing and projected accelerator searches. We will outline the relevant bounds on this coupling below.

\begin{figure*}[ht!]
    \centering
    \begin{subfigure}[t]{0.49\textwidth}
         \centering
         \includegraphics[width=\textwidth]{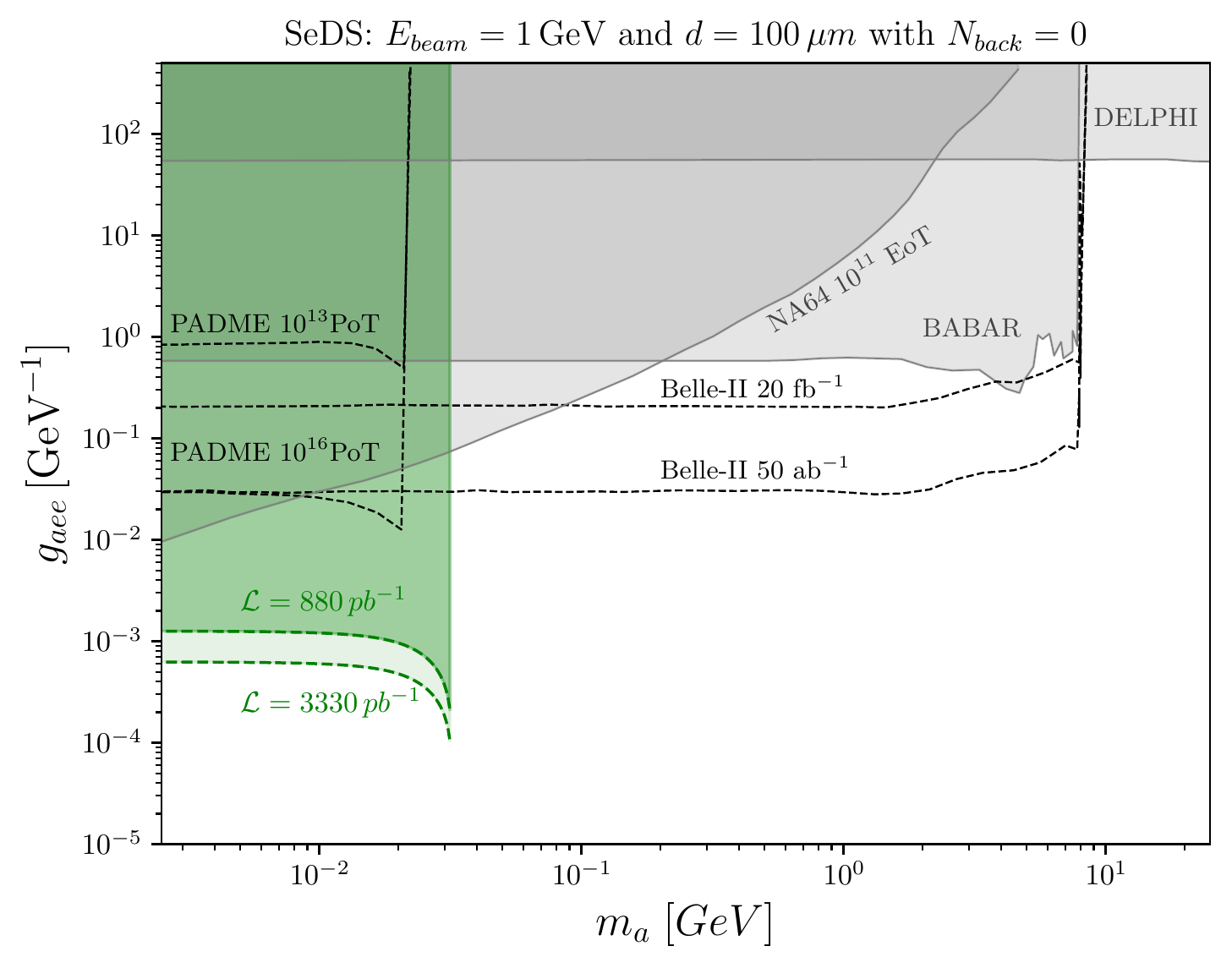}
         \caption{Sensitivity for $E_{beam}=1 \, \mbox{GeV}$.}
         \label{fig:sensitivity1gev}
     \end{subfigure}
     \hfill
     \begin{subfigure}[t]{0.49\textwidth}
         \centering
         \includegraphics[width=\textwidth]{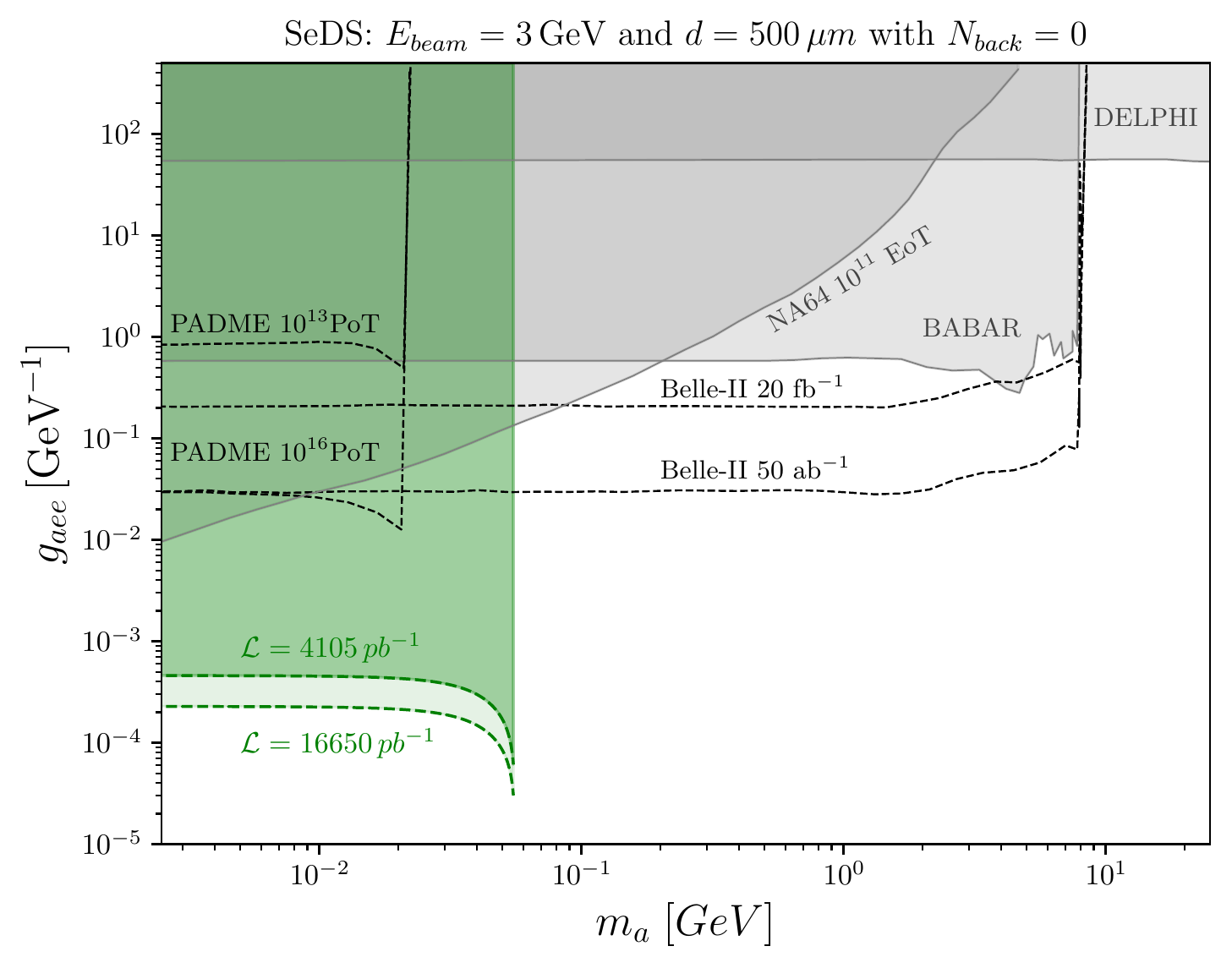}
         \caption{Sensitivity for $E_{beam}=3 \, \mbox{GeV}$.}
         \label{fig:sensitivity3gev}
     \end{subfigure}
    \caption{Current and projected bounds for the $g_{aee}$ coupling in ALP searches for the case where $g_{a\gamma \gamma}=0$. SeDS' sensitivity for $N_{back} \sim 0$ is presented in the green region, while the current and projected limits considered here were taken from \cite{Darme:2020sjf}, where the solid (dashed) gray (black) lines represent the current (projected) limits from DELPHI, Babar, Belle-II, NA64, and PADME.}
    \label{sensitivityplot}
\end{figure*}

\textbf{DELPHI-LEP}: The DELPHI experiment was a detector inside the Large Electron Positron (LEP) collider at CERN \cite{Alekseev:2001va}. A revision of the data from the DELPHI experiment on monophoton processes of the type $e^+ e^- \to X \gamma$ resulted in $g_{aee}<8\times 10 \,\mbox{GeV}^{-1}$.  The excluded region is shown in Fig. \ref{sensitivityplot}.

\textbf{BaBar}: BaBar is a $e^+e^-$ collider \cite{HARRISON199581}. The BaBar Collaboration analyzed the production of dark photon in the process $e^+ e^- \to A^\prime \gamma$ where $A^\prime$ represents the dark photon, decaying into $e^+ e^-,\mu^+ \mu^-$ \cite{Lees_2014}. Recasting this bound, we find $g_{aee} < 0.6 \,\mbox{GeV}^{-1}$ for
$1 \, \mbox{MeV} < m_a < 10 \, \mbox{GeV}$. This constraint is exhibited in Fig. \ref{sensitivityplot}.

\textbf{Belle-II}: The Belle-II experiment \cite{Kou_2019} has a similar setup to BaBar, being able to search for visible and invisible ALPs. The projected bounds coming from Belle-II are initially on $g_{a\gamma \gamma}$, but they can be translated to $g_{aee}$. In this work, we considered the Belle-II projections for the luminosity of 20 fb$^{-1}$ and 50 ab$^{-1}$. This high luminosity makes this projection produce the most compelling constraints covering most regions of parameter space, being the best probe for ALPs in the range of hundreds of MeV to 10 GeV. The projection for both luminosity configurations is given in Fig. \ref{sensitivityplot}. 

\textbf{NA64}: The NA64 experiment \cite{NA64:2018lsq} consists of a beam of electrons with an energy of 100 GeV dumped at a fixed target. The experiment goal is to search for Dark Photons via the process $e^- + Z \to e^- + Z + A^\prime$, where $A^\prime$ is the dark photon that can decay visibly and invisibly, being able to analyze data from the decays $A^\prime \to e^+ e^-,\gamma \gamma$. The results of NA64 were used to search for ALPs \cite{Banerjee_2020}. As no excess of events was observed, a bound was derived on the ALP-electron couplings, as shown in Fig. \ref{sensitivityplot}.   

\textbf{PADME}: The PADME experiment features a positron beam with variable energy of $200 - 550$ MeV directed at an active target. Measuring the energy of the final states, we can constrain the presence of invisible particles such as ALPs \cite{Raggi_2014}. This search achieved an integrated luminosity of $4\times10^{13}$ POT with no positive signal. A new run is planned with $4\times10^{16}$ POT \cite{Darme:2020sjf}. The current and projected bounds are exhibited in Fig. \ref{sensitivityplot}.  

\subsection{SM Background Processes}

Standard model processes can generate a signal that resembles $e^+e^-\to a\gamma$. However, it is crucial to note that in the framework we are considering, the ALP has a long decay length and therefore decays outside the detector, resulting in signal events characterized by a single photon in the electromagnetic calorimeter. By using the energy of the beam, we can calculate the ALP mass using Eq. (\ref{missing-mass-tech}).

The primary background processes arise from $\sigma(e^+e^-\to\gamma\gamma)=0.93$ mb, $\sigma(e^+Z\to e^+\gamma Z)=2.2\times 10^3$ mb, $\sigma(e^+e^-\to e^+e^-\gamma)=77$ mb, and $\sigma(e^+e^-\to\gamma\gamma\gamma)=0.02$ mb for a positron beam energy of $1$~GeV. Among them, the largest cross-section originates from photon bremsstrahlung. However, this background can be effectively reduced because the photon from bremsstrahlung is almost collinear with the beam and the charged particle will be guided to the spectrometer via the electric dipole. Besides, a positron-electron veto system to further control this background could also be installed to further control this background \cite{Oceano:2022rsh}.

Similarly, the background arising from $e^+e^-\gamma$ production can be managed. Although its cross section is not particularly large, the $\gamma\gamma\gamma$ production presents a challenging background. Due to the phase space, there is a lack of symmetry, which could result in a false signal in the calorimeter, where one high-energy photon is detected, and two low-energy photons would pass undetected. This scenario leads to no significant peak in $M_{miss}^2$. Consequently, it is challenging to reduce this background by applying cuts on the energy and angular distribution of the calorimeter. Notice that this case is different from the two-photon annihilation production, where there is symmetry in the direction of the photons. 

We want to emphasize to the reader that this work is a theoretical proposal for an ALP search using a 3 GeV positron accelerator called SeDS, which is planned to be constructed using parts of a partly decommissioned 1.37 GeV accelerator at the LNLS. Our objective is to estimate the projected sensitivity of such an accelerator without considering detector effects but to demonstrate the physics potential of what could be the first Latin-American dark sector detector. In the following section, we explain how we obtained the projected limits. 

\begin{figure*}[t]
    \centering
    \begin{subfigure}[t]{0.48\textwidth}
         \centering
         \includegraphics[width=\columnwidth]{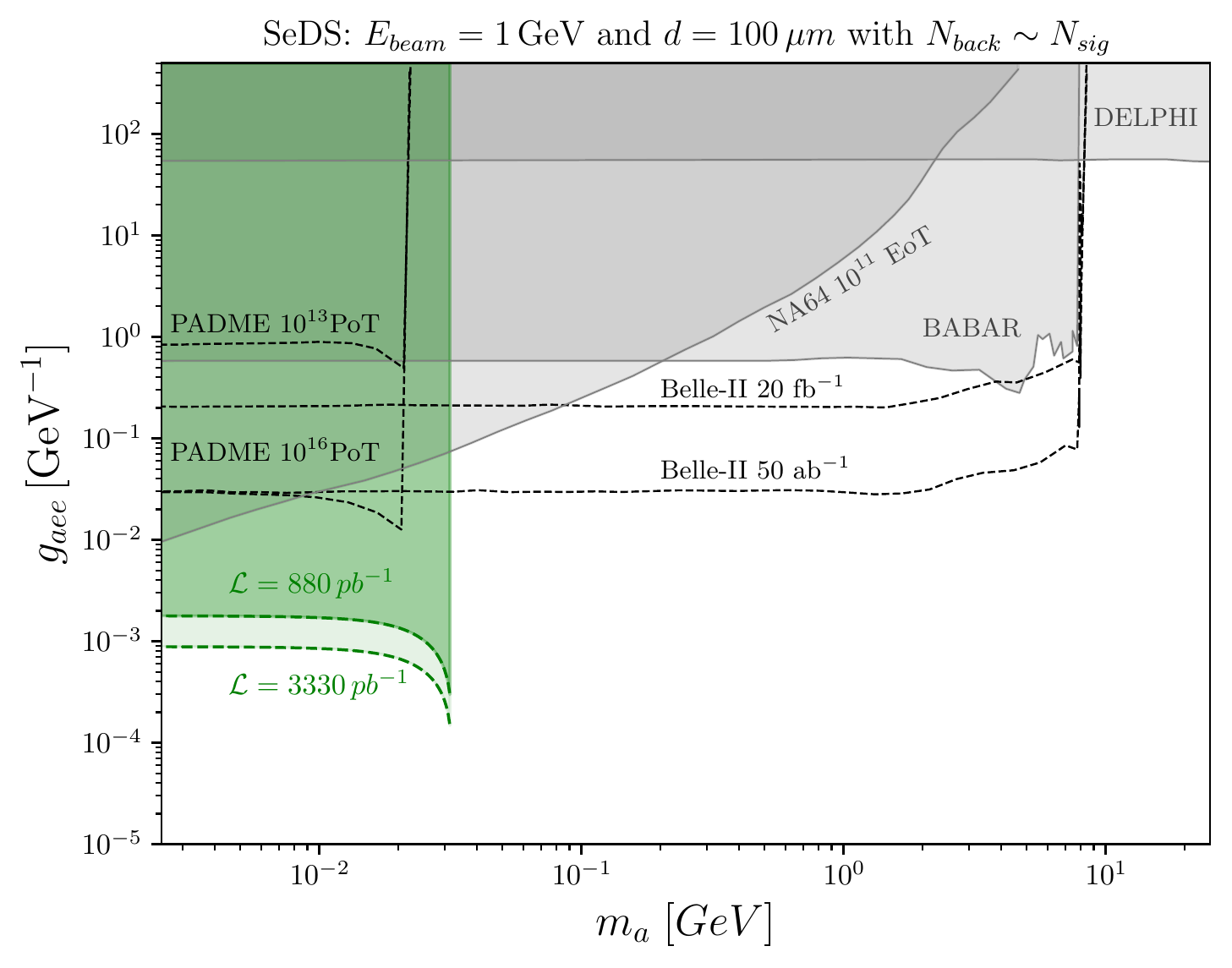}
         \caption{Sensitivity for $E_{beam}=1 \, \mbox{GeV}$.}
         \label{fig:sensitivity1gevback}
     \end{subfigure}
     \hfill
     \begin{subfigure}[t]{0.48\textwidth}
         \centering
         \includegraphics[width=\columnwidth]{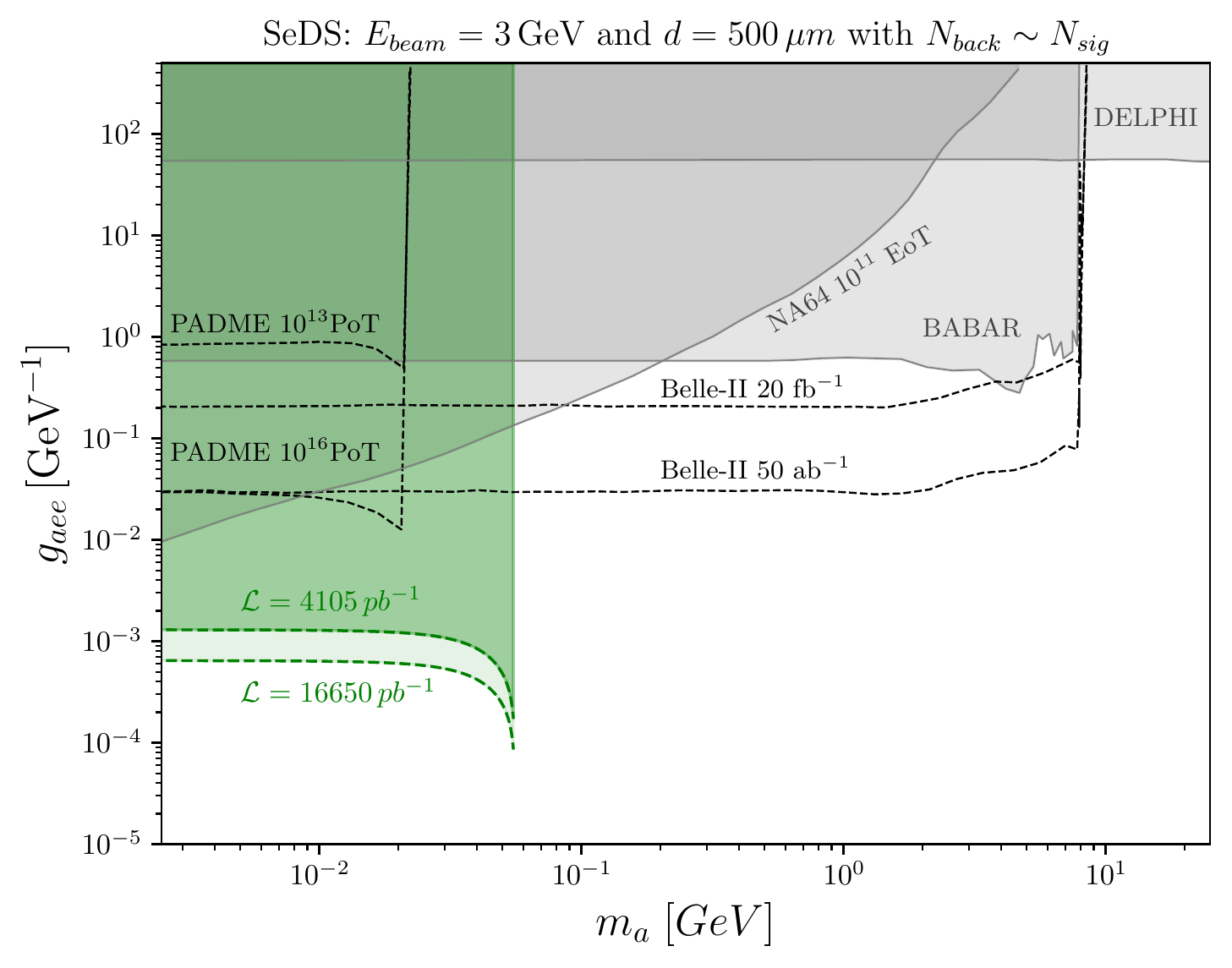}
         \caption{Sensitivity for $E_{beam}=3 \, \mbox{GeV}$.}
         \label{fig:sensitivity3gevback}
     \end{subfigure}
    \caption{Current and projected bounds for the $g_{aee}$ coupling in ALP searches for the case where $g_{a\gamma \gamma}=0$. SeDS' sensitivity for $N_{back} \sim N_{sig}$ is presented in the green region. The current and projected limits considered here were taken from \cite{Darme:2020sjf}, where the solid (dashed) gray (black) lines represent the current (projected) limits from DELPHI, Babar, Belle-II, NA64, and PADME.}
    \label{sensitivityplotback}
\end{figure*}

\begin{figure*}[t]
    \centering
    \begin{subfigure}[t]{0.49\textwidth}
         \centering
         \includegraphics[width=\columnwidth]{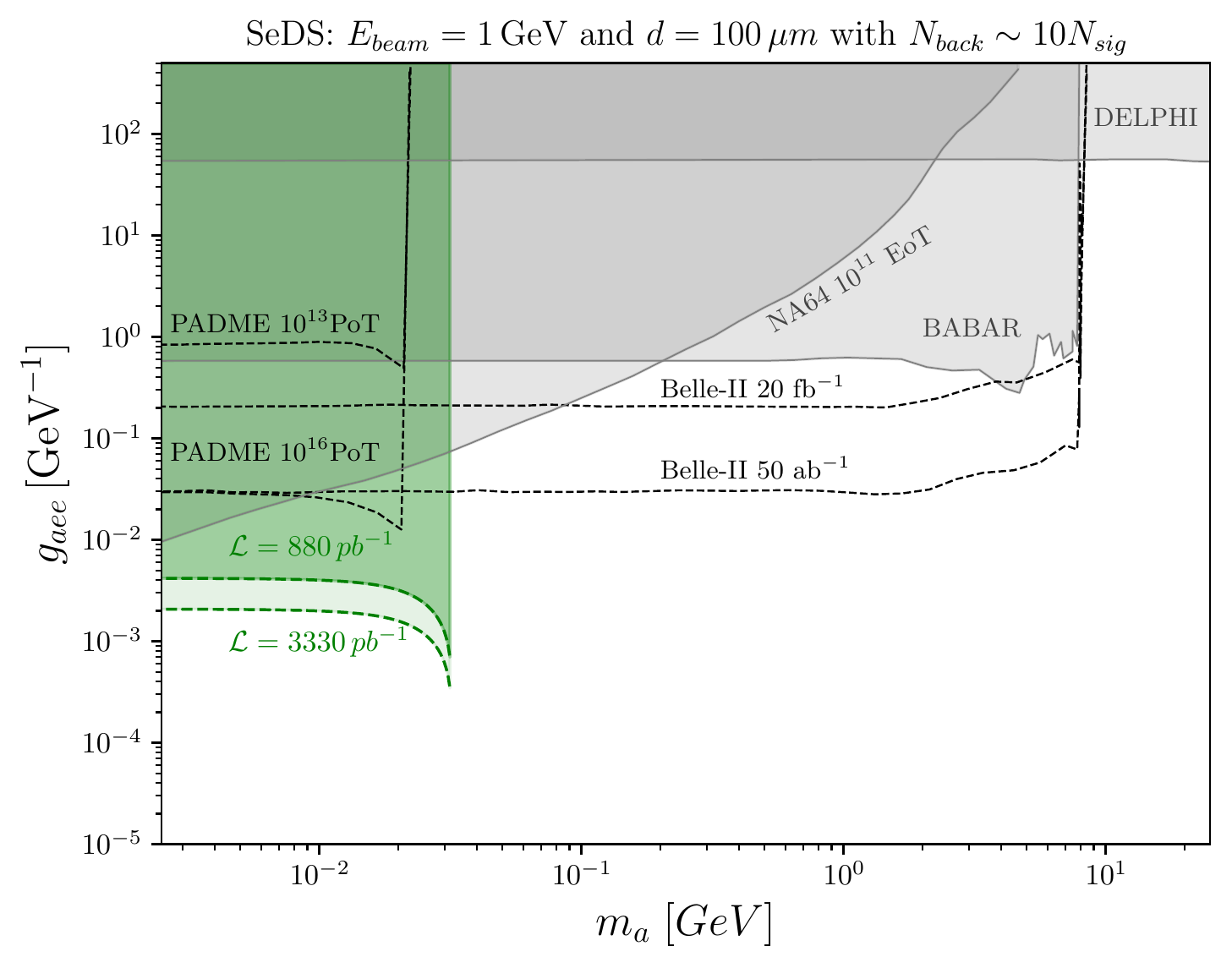}
         \caption{Sensitivity for $E_{beam}=1 \, \mbox{GeV}$.}
         \label{fig:sensitivity1gev10back}
     \end{subfigure}
     \hfill
     \begin{subfigure}[t]{0.49\textwidth}
         \centering
         \includegraphics[width=\columnwidth]{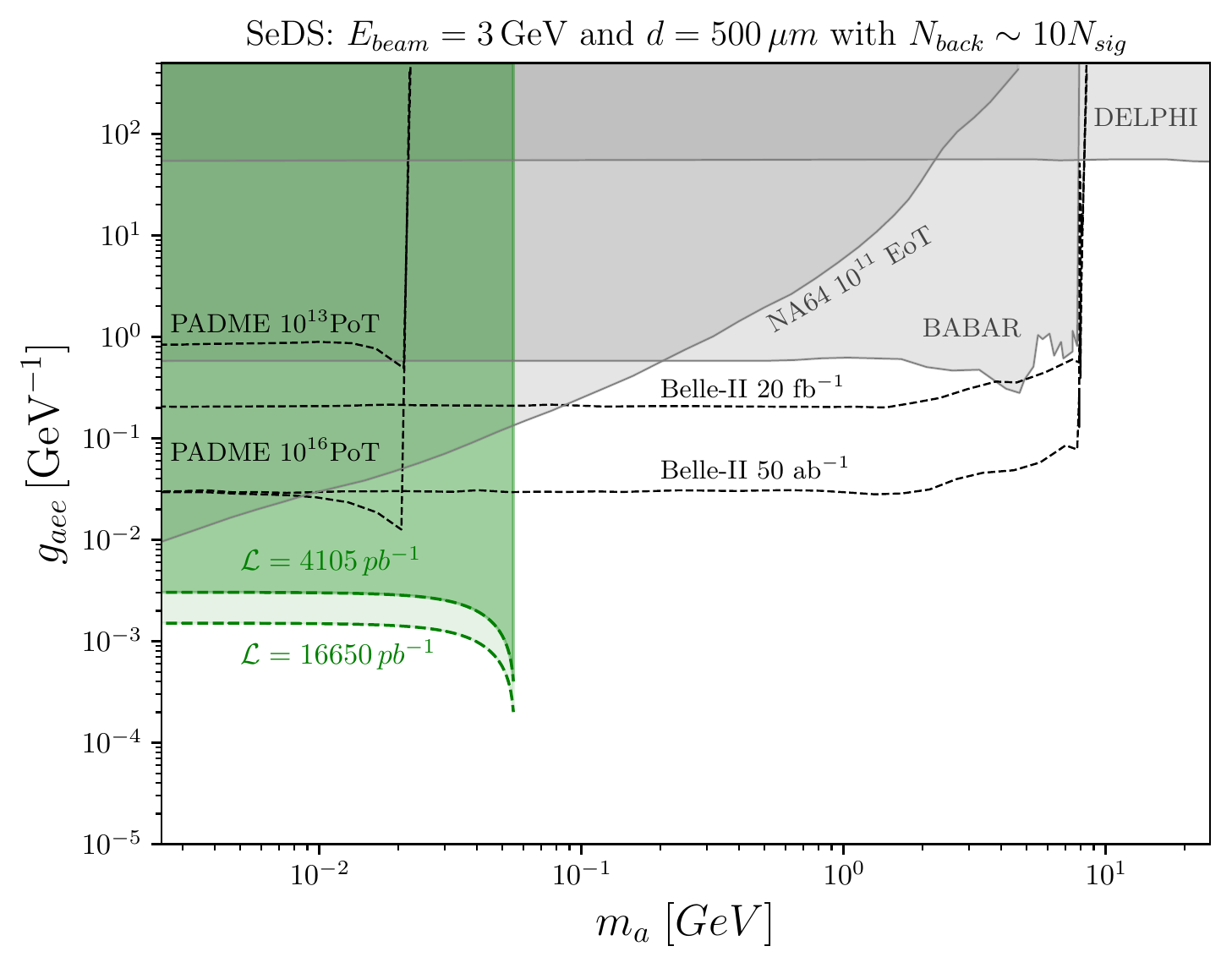}
         \caption{Sensitivity for $E_{beam}=3 \, \mbox{GeV}$.}
         \label{fig:sensitivity3gev10back}
     \end{subfigure}
    \caption{Same as Figures \ref{sensitivityplot} and \ref{sensitivityplotback}, but with $N_{back} \sim 10 N_{signal}$.}
    \label{sensitivityplot10back}
\end{figure*}

\subsection{Limits from SeDS}

An important quantity to assess the sensitivity of such an accelerator is the luminosity. The instantaneous luminosity of a positron beam impinging on a target is given by,
\begin{equation}\label{Lumi}
 L_{\text{inst}} = \frac{N_{\text{P.O.T}}}{\text{s}} N_{A} \frac{Z\rho d}{A} ,
\end{equation} where $N_{\text{P.O.T}}/\text{s}=10^{10}$ is the number of positrons on the target per second considering the properties of the decommissioned accelerator. Assuming a diamond target with a thickness of  $d=100 - 500\,\mu$m, we can calculate the instantaneous luminosity knowing that $N_A$ is the Avogadro number, $Z=6$ is the atomic number of the diamond target, $\rho=3,51$\,g/cm$^{3}$ is the density of diamond and $A = 12.01$\,g is the carbon's gram-molecular weight.

We consider different setups. One uses a 1 GeV positron beam, and the other uses a 3 GeV positron beam. With $E_{beam}=1$~GeV and $d=100\,\mu$m, we find 
$\mathcal{L}=821 \,\mbox{pb}^{-1}$ for 90 days of data taking, and $\mathcal{L}=3331 \,\mbox{pb}^{-1}$ for a one-year of data taking. For $E_{beam}=3$~GeV and $d=500\, \mu$m, we find 
$\mathcal{L}=4106 \,\mbox{pb}^{-1}$ for 90 days of data taking, and $\mathcal{L}=16650 \,\mbox{pb}^{-1}$ for a one-year of data taking.


Knowing the luminosity of the experiment, we can calculate the number of signal events and delimit the region of parameter space in which an ALP could be observed in the  $\{m_a,g_{aee}\}$ plane assuming $g_{a\gamma \gamma} =0$. 

In Fig. \ref{sensitivityplot}, we present the regions at 95\% C.L. assuming that the number of signal events is dominant over the number of background events ($N_{back} \sim 0$). In Fig. \ref{sensitivityplotback}, we consider the number of background events, which includes the irreducible background $(e^+e^- \to \gamma\gamma\gamma)$, to be similar to the number of signal events. Lastly, in Fig. \ref{sensitivityplot10back}, we assume the number of background events to be 10 times larger than the number of signal events. These figures demonstrate that SeDS is a promising probe for directly measuring ALP couplings to electrons, even in a very conservative scenario. It is important to note that these sensitivity projections do not take into account detector effects, but rather serve to demonstrate the physics potential of SeDS as the first Latin-American dark sector detector. In the following section, we explain in detail how we obtained these projected limits.

Fig. \ref{fig:sensitivity1gev} accounts the projected sensitivity of SeDS for the case $E_{beam} = 1 \,\mbox{GeV}$ and $d = 100 \, \mu\mbox{m}$. In this setup, SeDS probes ALP-electron coupling down to $g_{aee} \sim 6 \times 10^{-4} \, \mbox{GeV}^{-1}$ for masses up to $m_{a} \sim 32 \,\mbox{MeV}$. Fig. \ref{fig:sensitivity3gev} exhibits the sensitivity for $E_{beam}=3 \,\mbox{GeV}$ and $d = 500 \, \mu\mbox{m}$. In the latter setup, SeDS could probe ALP-electron coupling down to $g_{aee} \sim 2.3 \times 10^{-4} \, \mbox{GeV}^{-1}$ covering larger ALP masses up to $m_{a}\approx 55 \, \mbox{MeV}$ due to the larger center of mass energy. It is noteworthy that the higher luminosity and center of mass energy of the latter configuration can achieve better results than the first one. In both cases, SeDS presents a promising prospect in the search for an electrophilic axion-like particle and projects better results than other experiments of its kind.

However, it is important to note that the sensitivity presented in Fig. \ref{sensitivityplot} is the result of an optimistic analysis in which the experiment's background can be reduced as much as possible, ensuring that the signal is considerably higher than the background. To present a more realistic scenario of SeDS's sensitivity for the electron-mediated channel, Fig. \ref{sensitivityplotback} shows the case where $N_{back} \sim N_{signal}$.

The first scenario considered is the setup with $E_{beam} = 1 \,\mbox{GeV}$ and $d = 100 \, \mu\mbox{m}$, presented in Fig. \ref{fig:sensitivity1gevback}. In this case, SeDS can probe the ALP-electron coupling for values of $g_{aee} > 9 \times 10^{-4} \,\mbox{GeV}^{-1}$ with $\mathcal{L} = 3300 \,\mbox{pb}^{-1}$. In the second experimental setup where $E_{beam} = 3 \,\mbox{GeV}$ and $d = 500 \, \mu\mbox{m}$, SeDS can reach an ALP-electron coupling of $g_{aee} > 6.5 \times 10^{-4} \,\mbox{GeV}^{-1}$. As expected, SeDS's sensitivity is reduced when there is a non-zero background. Nevertheless, SeDS's projected sensitivity is still an improvement compared to the results of similar experiments. 

In a similar vein, we consider a very conservative scenario, we repeat the exercise above assuming the number of background events to be ten times larger than the signal events, $N_{sig}=10\, N_{back}$, in Fig. \ref{fig:sensitivity1gev10back} and Fig. \ref{fig:sensitivity3gev10back}. Even so, SeDS covers an unexplored region of parameter space.

\section{\label{sec13} Conclusion}

In this work, we have proposed the first direct search for axion-like particles at the LNLS using a detector named SeDS, which consists of a $1-3$~GeV accelerator impinging on a target planned to be constructed using subsystems of UVX, a $1.37$~GeV accelerator which has been recently decommissioned. We have concluded that SeDS could place competitive bounds on the coupling of axion-like particle coupling to electrons. In particular, in a conservative scenario, we have shown that SeDS could surpass current and projected experiments in the mass range of 1-30 MeV using a $1$~GeV positron beam. In an optimistic configuration with a $3$~GeV positron beam, SeDS could potentially probe axion-like particles for masses up to 55 MeV and outperform current and  projected experiments of the kind, probing $g_{aee}$ down to $3 \times 10^{-4} \, \mbox{GeV}^{-1}$. 

\acknowledgments
We would like to thank Harry Westfahl, Daniel Tavares, and Narcizo Neto from the LNLS for the hospitality and providing useful information about the accelerator.
\begin{itemize}
\item {\bf Funding:} Simons Foundation (Award Number:1023171-RC), FAPESP Grant 2018/25225-9,
2021/01089-1, 2023/01197-4, ICTP-SAIFR FAPESP Grants 2021/14335-0, CNPq Grant 307130/2021-5, ANID-Programa Milenio-code ICN2019\_044, CAPES grant 88887.485509/2020-00, ANID-Programa Milenio-code ICN2019\_044, ANID PIA/APOYO AFB180002 (Chile), ANID/CONICYT FONDECYT Regular 1221463 and FONDECYT Regular 1210131. Y.V. expresses gratitude to São Paulo Research Foundation (FAPESP) under Grant No. 2018/25225-9 and 2023/01197-4.
\item No conflict of interest to be reported.
\item {\bf Authors' contributions:} Álvaro S. de Jesus and Farinaldo S. Queiroz conceived the study, contributed to the writing and mathematical computations of the manuscript. Paola Arias, Claudio O. Dib, Sergey Kuleshov, Venelin Kozhuharov, Manfred Lindner, Liu Lin, Yoxara Villamizar, Lucia Angel, and Ricardo C. Silva contributed to the writing and phenomenology of the work. 

\end{itemize}

\noindent

\bibliographystyle{unsrt}
\bibliography{bibliography}

\end{document}